\documentclass[12pt,letter]{article}

\usepackage{graphics,graphicx,multirow,fullpage,setspace,wrapfig,multicol,enumerate}
\usepackage[bf,font=footnotesize,textfont=it]{caption}
\usepackage[version=3]{mhchem}
\usepackage{cite,titlesec}
\usepackage{hyperref}
\newlength\tindent
\usepackage[all]{nowidow}
\usepackage{color}
\usepackage{floatrow}
\usepackage[margin=1.0in]{geometry}

\titleformat*{\section}{\large\bfseries}

\begin{document}

\thispagestyle{empty}


\newpage
\onecolumn

\begin{figure}
    \centering
    \includegraphics[width=\textwidth,height=190pt]{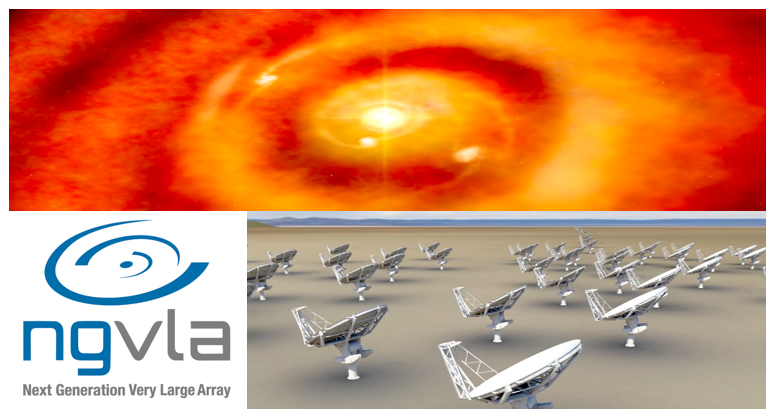}
\end{figure}

\begin{center}
\Large
\textsc{Witnessing Planetary Systems in the Making \\ with the Next Generation Very Large Array}\\
\normalsize
\end{center}

\begin{center}

    Luca Ricci\\
    \vspace{0.5em}
    \emph{California State University Northridge, Northridge, CA 91130 \\
    Jet Propulsion Laboratory, Pasadena, CA 91109\\
    E-mail: luca.ricci@csun.edu\\
    }
    
   \vspace{0.7em}
    
    With the Support Of\\\vspace{1em}
    
    \begin{minipage}{0.85\textwidth}
    
    \begin{tabular}{l l}

    Andrea Isella                   & \emph{Rice University}\\
    Sean M. Andrews                 & \emph{Harvard-Smithsonian Center for Astrophysics}\\
    Tilman Birnstiel                & \emph{Universit\"{a}ts-Sternwarte Munich}\\
    Jeffrey N. Cuzzi                & \emph{NASA Ames Research Center}\\
    Gennaro D'Angelo                & \emph{Los Alamos National Laboratory}\\
    Ruobing Dong                  & \emph{Steward Observatory}\\
    Anne Dutrey                & \emph{Laboratoire d'Astrophysique de Bordeaux}\\
    Barbara Ercolano                & \emph{Universit\"{a}ts-Sternwarte Munich}\\
    Paul R. Estrada                & \emph{SETI Institute}\\
    Mario Flock                & \emph{Jet Propulsion Laboratory}\\
    Hui Li                & \emph{Los Alamos National Laboratory}\\
    Shang-Fei Liu                & \emph{Rice University}\\
    Wladimir Lyra                & \emph{California State University Northridge}\\
    Karin \"{O}berg                & \emph{Harvard-Smithsonian Center for Astrophysics}\\
    Satoshi Okuzumi                & \emph{Tokyo Institute of Technology}\\
    Laura Perez                & \emph{Universidad de Chile}\\
    Neal Turner                & \emph{Jet Propulsion Laboratory}\\
    Nienke van der Marel                & \emph{Herzberg Astronomy and Astrophysics}\\
    David Wilner                    & \emph{Harvard-Smithsonian Center for Astrophysics}\\
    Andrew N. Youdin                & \emph{University of Arizona}\\
    Zhaohuan Zhu                & \emph{University of Nevada, Las Vegas}\\

    \end{tabular}
    
    \end{minipage}
    
\end{center}

\newpage

\normalsize
\subsection*{Executive Summary}

The discovery of thousands of exoplanets over the last couple of decades has shown that the birth of planets is a very efficient process in nature \cite{Burke:2015}. 
Theories invoke a multitude of mechanisms to describe the assembly of planets in the disks around pre-main-sequence stars, but observational constraints have been sparse on account of insufficient sensitivity and resolution. 
Understanding how planets form and interact with their parental disk is crucial also to illuminate the main characteristics of a large portion of the full population of planets that is inaccessible to current and near-future observations.

\vspace{2mm}

Two key issues hamper our understanding of planet formation. 
The first issue is that, since the very first detection of an exoplanet around a Solar-like star \cite{Mayor:1995}, the discovery of several `hot Jupiters' has indicated that the orbits of planets can significantly vary right after their formation. 
In particular, newborn planets can easily radially migrate either inward or outward through exchange of angular momentum 
with the parent circumstellar material. This means that the orbital parameters of mature exoplanets likely do not correspond 
to the \textit{initial} location of their formation. Knowing where planets form in the disk is crucial to constrain the dynamical, physical and chemical history of planetary systems.

\vspace{2mm}

Secondly, our current understanding of planet formation posits that planets form through the agglomeration of small dust particles into much larger `planetesimals', which are massive enough to gravitationally attract other solids in the disk. Yet, the formation of these planetesimals is one of the most critical problems for theories of planet formation. In the simplest assumption of a gas-rich disk with density and temperature decreasing with distance from the star, small solids radially drift inward as a consequence of the aerodynamic drag by the gas orbiting at sub-Keplerian speeds \cite{Weidenschilling:1977,Johansen:2014}. Models of the evolution of disks solids have calculated radial drift timescales which are too short to form planetesimals \cite{Testi:2014}. 

\vspace{2mm}

Since the past few years, the astronomy community has
initiated discussion for a future millimeter/radio
array optimized for the imaging of thermal emission to milliarcsecond
(mas) scales, corresponding to 1 au or less in nearby disks, that will open a new discovery space for
studies of proto-planetary disks.
Although the final design of the array has yet to be established,
this Next Generation Very Large Array (ngVLA) is currently
envisioned to observe at frequencies of $\sim 1.2 - 116$ GHz and to
include $\sim$10$\times$ the collecting area of the VLA and ALMA and
$\sim$10$\times$ longer baselines.

\vspace{2mm}

With these capabilities, the ngVLA will transform our understanding of planet formation by:

\begin{itemize}
    \item probing the locations and measuring the masses of forming giants and Super-Earths in the planet-forming regions of nearby young disks;
    \item detecting and characterizing the local concentrations of small solids in the disk which trace the birth sites of planetesimals.
    
\end{itemize}

\subsection*{Finding and Characterizing Newborn Planets in Young Disks}

Only recently, infrared and (sub-)millimeter telescopes have achieved the angular resolution
required to spatially resolve the inner regions of nearby proto-planetary disks, down to distances of $\sim 10 - 20$ au from the central star. 
Observations at sub-mm and longer wavelengths probe the emission from solids in the disk midplane, and multi-wavelength observations can be used to constrain the sizes of the emitting solids \cite{Ricci:2010}. 

\vspace{2mm}

Recent high angular resolution observations at these wavelengths 
resulted in the discovery
of morphological features, such as rings, spirals, and crescents, in the distribution of circumstellar gas and dust with characteristic sizes larger than 20 au \cite{Casassus:2013,ALMA:2015,Andrews:2016,Isella:2016,Tang:2017}. 
These structures are suggestive of gravitational perturbations of yet unseen newborn giant planets \cite{Bryden:1999}, and provide a powerful tool to measure planet masses and orbital radii, and investigate how forming planets interact with the circumstellar material \cite{Jin:2016}. 

\vspace{2mm}

The same observations also show that, at sub-mm wavelengths, the dust continuum emission arising within 10$-$30 au from the star is often optically thick \cite{Carrasco:2016}. The large optical depth prevents us from measuring the dust density and, therefore, image planet-driven density perturbations. In practice, observations with instruments like ALMA, NOEMA or the SMA, cannot probe planets orbiting in these innermost regions, unless the planets are massive enough to carve very deep, optically thin, gaps in the dust distribution. Therefore, observations with current telescopes miss the bulk population of forming planets, which have masses way below the mass of Saturn and orbit at less than 10 au from the host star \cite{Burke:2015}. 

\vspace{2mm}

A natural solution to this problem consists in imaging proto-planetary disks with spatial resolutions of about 1 au at wavelengths longer than 1 mm, where the dust continuum emission from the innermost disk regions is optically thin, but still bright enough to be detected. These wavelengths are covered by the Karl G. Jansky Very Large Array (VLA), the Australia Telescope Compact Array (ATCA), and in the near future by ALMA Band 1 and 2, which, however, lack the angular resolution and sensitivity to efficiently search for signatures of young planets. 

\vspace{2mm}

These capabilities would instead be achievable with the ngVLA. Thanks to its large collecting area and resolving power, the ngVLA will provide for the first time imaging of the disk thermal emission down to mas scales at wavelengts longer than 1 mm, which will be transformational for studies of protoplanetary disks. 

\vspace{2mm}

The potential of the ngVLA to observe substructures in nearby young disks caused by the gravitational interaction between disk material and planets between 1 and 5 au from the star was the subject of a recent investigation \cite{Ricci:2018}. It was found that ngVLA observations with an angular resolution of 5 mas at 3 mm can detect and characterize 
structures generated by planets with masses as low as $5~M_{\oplus}$ surrounding young Solar-like stars in the closest star forming regions. 

\vspace{2mm}

\begin{figure}
    \centering
    \includegraphics[scale=0.42]{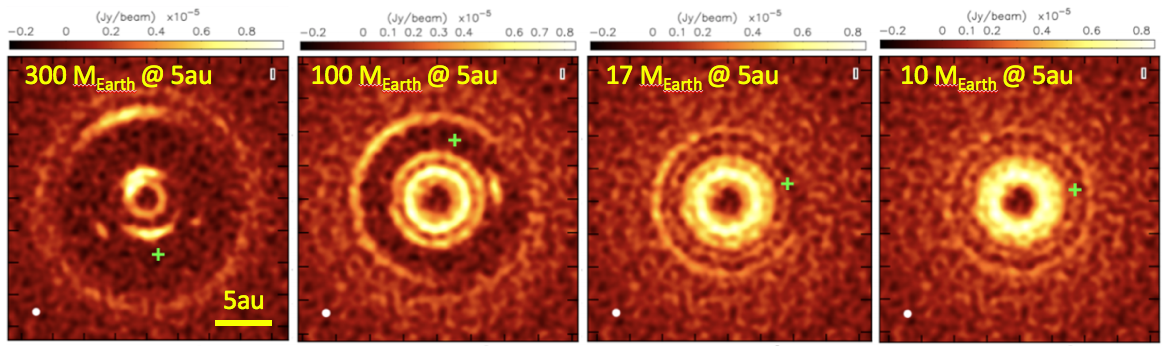}
    \caption{Simulations of ngVLA observations at a wavelength of 3 mm of a disk with planets at 5 au from the central star. The planet masses are labeled in each panel. The white ellipses show the size of the synthesized beam, i.e. 5 milliarcsec, corresponding to 0.7 au at the assumed distance of 140 pc \rm{\cite{Ricci:2018}}.}
    \label{fig:ngvla_planets}
\end{figure}

Figure~\ref{fig:ngvla_planets} shows that the ngVLA can detect and spatially resolve strong azimuthal asymmetries in the disk generated by the interaction with planets as massive as Saturn ($\approx$ 100~$M_{\oplus}$) or heavier, as well as more symmetric structures and narrower gaps due to lower mass planets. 
The ngVLA will also be able to detect, and separate from the emission of the circumstellar disks, the emission from \textit{circumplanetary} disks that have properties similar to those predicted by models of formation of the natural satellites around Jupiter and Saturn \cite{Zhu:2017}.  
Moreover, given its very high astrometric precision, the ngVLA can measure the orbit of these features on monthly timescales \cite{Ricci:2018}.  

\vspace{2mm}

Thanks to its exquisite sensitivity and angular resolution, the 
ngVLA is capable of indirectly detect the presence of super-Earths planets, and measure their birth orbital radii, in several dozens of disks within 200 pc from the Sun \cite{Ricci:2018}.
The unprecedented ngVLA capabilities would also allow to extend the search for giant exoplanets in the act of forming to hundreds of disks as far as 700 pc from the Sun. 
The statistical comparison between the properties constrained for the population of young planets still embedded in their parental
disks with those obtained for more mature exoplanets by past
and future facilities (e.g., Kepler, JWST)  will provide key observational
tests to the models of formation and evolution of planetary systems.

\vspace{-2mm}

\subsection*{Detecting the Birth Sites of Planetesimals}

As described in the first section, our understanding of the formation of planetesimals, and therefore of planets, is hampered by the inward radial drift of small solids expected in a smooth gas-rich disk.
This inward radial drift can be slowed down, and even stopped, if there are local \textit{over-densities} in the gas, which are capable of trapping particles. In these regions the local dust-to-gas mass ratio may increase up to the point at which the dusty layer becomes unstable and fragments. These fragments might eventually gravitationally collapse and form planetesimals. The origins of these over-densities are likely dynamic or magneto-hydrodynamic instabilities \cite{Lovelace:1999,Li:2000,Nelson:2013,Klahr:2014,Lyra:2014,Flock:2015}. Therefore, a natural prediction of these models is the presence of strong local accumulations of small solids in the disk \cite{Johansen:2014}. 

\vspace{2mm}

\begin{figure}
    \centering
    \includegraphics[scale=0.5]{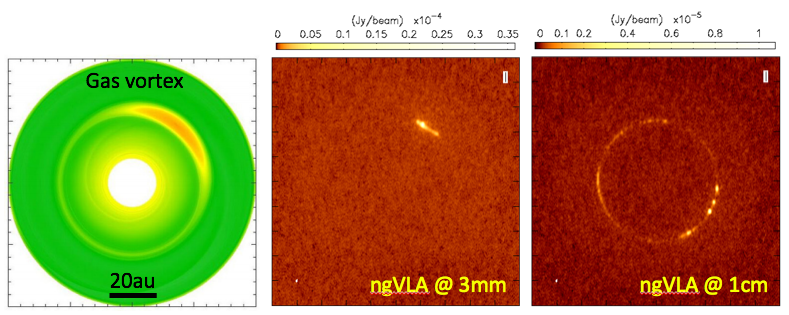}
    \caption{Simulations of ngVLA observations for a disk with a gas vortex. The gas density map for the disk model with a vortex in the north-west side of the disk is shown on the left panel. The center and right panels show the ngVLA observations at wavelenghts of 3 mm and 1 cm, respectively. The sizes of the ngVLA synthesized beams are shown in the lower left corner of the center and right panets \rm{\cite{Barge:2017}}.}
    \label{fig:ngvla_vortex}
\end{figure}

In the last decade, sub-mm interferometers like the SMA, CARMA and the PdBI, and, more recently ALMA, have found striking evidence for  local accumulations of small solids at distances $>$ 30 - 50 au from the star \cite{Brown:2009,Isella:2013,vanderMarel:2013,Casassus:2013}, as well as for a local increase of the dust-to-gas mass ratio in those regions \cite{Perez:2014,Boehler:2018}. 
These observations support the existence of some of the instabilities invoked by theory to explain the origin of planetesimals. However, due to the limited resolution (in the Solar System, 30 - 50 au correspond to the Kuiper Belt) they only probe the outskirts of the planet-forming region of these disks.  


\vspace{2mm}

Furthermore, models predict that solids with sizes of $\sim$ 1 - 10 mm should be more concentrated toward the center of local gas over-densities than smaller dust grains. 
Solids with these sizes can be best traced by mapping the dust thermal emission at wavelengths longer than 1 mm. 
The wavelength range at which the ngVLA operates, coupled with high sensitivity and angular resolution, is optimal  
to unravel the birth sites of planetesimals in the planet-forming regions of disks. By investigating the morphology of the local accumulations of small solids down to scales of about 1 au, the ngVLA observations have the potential to differentiate between different physical processes that have been proposed to trigger the concentration of solids.

\vspace{2mm}

One example of such investigations is shown in Figure~\ref{fig:ngvla_vortex}, which presents the results of ngVLA simulated observations for a nearby disk with solids concentrated by a gas vortex \cite{Barge:2017}. The ngVLA observations at 3 mm trace the distribution of mm-sized particles which are strongly concentrated in a narrow arc structure by the gas vortex. In these models, larger cm-sized pebbles leave the location of the vortex along the azimuthal direction. This results into a ring-like morphology of the emission with significant azimuthal asymmetry observable at cm wavelengths. The densest clumps shown here contain several Earth masses in small solids, which can be converted into large planetesimals through their gravitational collapse.

\vspace{-3mm}

\subsection*{Summary}


The ngVLA will measure the planet initial mass function down to a
mass of $5 - 10$ Earth masses and unveil the formation of planetary systems similar to our own Solar System by probing the
presence of planets on orbital radii as small as 0.5 au at the distance of 140 pc. The investigation of small substructures in the distribution of small solids in young disks will unveil the physical mechanisms that drive the formation of planetesimals, the building blocks of planets. The statistical comparison between the properties of young planets with those obtained for more mature exoplanets by past
and future facilities (e.g., Kepler, JWST)  will provide key observational
tests to the models of formation and evolution of planetary systems.

\footnotesize
\twocolumn
\bibliography{bibliography}
\bibliographystyle{jobprops}

\end{document}